\documentclass[aps,pre,twocolumn,float,amsmath]{revtex4}
\usepackage[dvips]{color}
\usepackage[normalem]{ulem} 
\definecolor{g-blue}{rgb}{0.83,0.95,1}
\definecolor{g-yellow}{rgb}{1,1,0.7}
\definecolor{g-green}{rgb}{0.9,1,0.9}
\definecolor{green}{rgb}{0,0.6,0}
\definecolor{cyan}{rgb}{0,0.7,0.7}
\definecolor{black}{rgb}{0,0,0}
\definecolor{grey}{rgb}{0.4 ,0.4 ,0.4 }

\usepackage{bm}
\usepackage{amssymb}
\usepackage{amsfonts}
\usepackage{amsmath}
 \usepackage{graphicx}
  \usepackage{amsmath,bm,epsfig}
\def \ed {\end{document}}
\def\Fbox#1{\vskip1ex\hbox to 8.5cm{\hfil\fboxsep0.3cm\fbox{%
  \parbox{8.0cm}{#1}}\hfil}\vskip1ex\noindent}  %%  {TEXT} in BOX

\newcommand{\eq}[1]{(\ref{#1})}%%  requires \eq{label}
\newcommand{\Eq}[1]{Eq.\,(\ref{#1})}%%  requires \eq{label}
\newcommand{\Eqs}[1]{Eqs.\,(\ref{#1})}%%  requires \eq{label}
\newcommand{\Fig}[1]{Fig.\,\ref{#1}}%%  requires \Fef{label}
\def\be{\begin{equation}}\def\ee{\end{equation}}
\def\bea{\begin{eqnarray}}\def\eea{\end{eqnarray}}
\def\bse{\begin{subequations}}\def\ese{\end{subequations}}
\newcommand{\BE}[1]{\begin{equation}\label{#1}}
\newcommand{\BEA}[1]{\begin{eqnarray}\label{#1}}
\newcommand{\BSE}[1]{\begin{subequations}\label{#1}}

\let \nn  \nonumber

\let \= \equiv \let\*\cdot \let\~\widetilde \let\^\widehat \let\-\overline

  \def\1{\bm1} 

\def\<{\left\langle}    \def\>{\right\rangle}
\def\({\left(}          \def\){\right)}
 \def \[ {\left [} \def \] {\right ]}

\newcommand{\B}[1]{{\bm{#1}}}%% Bold Roman & Greek Lower & Upper Case
\newcommand{\C}[1]{{\mathcal{#1}}}    %%   Calligrapfic Upper case
\newcommand{\BC}[1]{\bm{\mathcal{#1}}}%% Bold Calligrapfic Upper case
\renewcommand{\sb}[1]{_{\text {#1}}}  %% sub-   for lower case
\def\He4 {$^4$He~}

\begin{document}

\title{Mechanical momentum transfer in wall-bounded superfluid turbulence}
\author{D. Khomenko, V.S. L'vov, A.  Pomyalov and I.  Procaccia }
\affiliation{Department of Chemical Physics,  Weizmann Institute  of Science, Rehovot 76100, Israel }

\begin{abstract}

 In classical turbulence the kinematic viscosity $\nu$ is involved in two phenomena. The first is the energy dissipation and  the second is the mechanical momentum flux toward the wall.  In superfluid turbulence the mechanism of energy dissipation is different, and it is determined by an effective viscosity which was introduced by Vinen and is denoted as $\nu'$. In this paper we show that in superfluid turbulence the transfer of mechanical momentum to the wall is caused by the presence of a quantum vortex tangle, giving rise to another effective ``momentum'' viscosity that we will denote as $\nu\sb {m}(T)$. The temperature dependence of the second effective viscosity is markedly different from  Vinen's effective viscosity $\nu'(T)$. We show that the notion of vortex-tension force, playing an important role in the theory of quantum turbulence, can be understood as the gradient of the Reynolds stress tensor which is in fact determined by the second newly defined  kinematic viscosity $\nu\sb {m}(T)$.
\end{abstract}

\maketitle

\section{Introduction}
\label{intro}
 Below the Bose-Einstein condensation temperature $T_\lambda\approx 2.18\,$K, liquid \He4 begins to gain a component that can be modeled as a quantum inviscid superfluid\,\cite{Donnely,DB98}. At intermediate temperatures below $T_\lambda$ the relative
 concentrations of normal fluid and superfluid components change in favor of the superfluid component, until eventually
 at sufficiently low temperature \He4 becomes inviscid. In addition to the inviscid nature of the
 superfluid, the vorticity is then carried by vortex-line singularities of fixed circulation $\kappa= h/M$, where $h$ is Planck's constant  and $M$ is the mass of the \He4 atom\,\cite{Feynman}.  These vortex lines have
 a core radius $a_0\approx  10^{-8}\,$cm, comparable to the inter-atomic distance. In generic turbulent state, these vortex lines appear as a complex tangle with a typical inter-vortex distance  $\ell\sim 10^{-4}\div 10^{-2}\,$cm\,\cite{Vinen}.
As long as the superfluid component coexists with a normal (viscous) fluid component, the large scale (much larger than $\ell$) hydrodynamic properties of \He4 are well described by the well known two-fluid model in which the two said components are coupled to each other. The coupling is modeled by the so-called ``mutual friction force", which is mediated by the tangle of quantized vortices\,\cite{Donnely,Vinen,HallVinen56,Vinen58,BK61}.

The aim of the present paper is to point out an important difference between classical and quantum turbulent flows.  The hydrodynamics of classical fluids is described by the Navier-Stokes equations, in which the viscosity term  is responsible for two different physical phenomena. The first is the damping of the kinetic energy of the flow, and it operates also in
 homogeneous isotropic flows. The second is a crucial ingredient of wall bounded flows, in which the friction between fluid layers with different {\em mean} velocities leads to the transfer of linear momentum  toward regions with smaller mean velocity\,\cite{Frisch,Pope}. The presence of an inviscid superfluid component in quantum turbulence requires a
 new outlooks on these related but distinct phenomena. It was understood that energy dissipation in the superfluid case is
caused fundamentally by the existence of the tangle of quantized vortex lines.
 The work of Vinen\,\cite{Vinen} and later studies\,\cite{Vinen2,SS,Nemir,WIS-2015} showed that the rate of  kinetic energy dissipation can be modeled well by the introduction of an effective ``energy" viscosity, which is traditionally denoted as $\nu'$.

Of course, also superfluid turbulence is usually neither homogeneous nor isotropic, being wall bounded as well. The superfluid velocity fluctuations caused by the dynamics of the tangle of quantized vortices lead to a momentum flux towards the wall. We show here that this momentum transfer can be well modeled by the introduction of a second effective ``momentum" viscosity $\nu\sb m $, different from the  effective energy viscosity $\nu'$. In particular we will show that the two
viscosities have different temperature dependence. The nature of the second effective viscosity will be clarified in this paper using two parallel strategies.  The first is analogous to classical hydrodynamics requiring the analysis of the Reynolds-stress tensor which in the present case is determined
by the velocity fluctuations created by the quasi-random vortex tangle. The second strategy is in line with the current description of quantum turbulence, using the notion of  the ``vortex tension force" in the presence of the vortex tangle. We demonstrate that these two strategies lead to the same result.  The vortex tension force, which is current in the superfluid turbulence literature, is analogous to the gradient of the Reynolds stress tensor, which is more familiar in the classical turbulence literature.

In Sect.~\ref{momflux} we introduce the notion of the effective momentum viscosity and propose a closure to
estimate it in terms of fundamental quantities. Since closure procedures are never rigorous, we
present in Sect.~\ref{numsim} numerical simulations to support and confirm the predictions of the
previous section. The concept of ``vortex tension force" is explained in Sect.~\ref{tense} where it
is shown that it is understandable as the gradient of the Reynolds stress that is caused by the
quantized vortex tangle. The presented analysis is supported again by numerical simulations. Finally, in
Sect.~\ref{summary} we offer a summary and conclusions.

\section{Viscosity and the turbulent momentum flux}
\label{momflux}
\subsection{Reminder from classical turbulence}

Consider a classical turbulent fluid flow in a channel geometry with the mean velocity in the $x$-direction and with $y$ being the wall normal direction. The mean velocity has only one component denoted as $V_x(y)$ and the turbulent velocity can
be decomposed into the mean plus fluctuations $\B{\tilde v}$,
\begin{equation}
\B v(\B r)= V_x(y) +\B{\tilde v}(\B r) \ ,
\end{equation}
The momentum flux towards the wall $\Pi_{xy}$ is given exactly by the sum of two contributions, the mean shear $S(y)$
and the so called Reynolds stress $W(y)$,
\begin{eqnarray}\label{flux}
\Pi_{xy}(y)&=& \nu\, S(y) + W(y)\,, \\ \nn
S(y)&=&\frac{\partial V_x}{\partial y}\,,\quad  W(y)=\< \~v_x \~v_y\> \ .
\end{eqnarray}
Here $\<...\>$ stand for the appropriate average (either over time or over an ensemble).
This is as far as one can go exactly.
To estimate the Reynolds stress we follow an old idea of Boussinesq\,\cite{2}, who suggested that in wall bounded flows $W(y)\propto S(y)$:
 \begin{equation}\label{est}
\< \~v_x \~v_y\>\simeq \nu_{\sb T} S(y)\ ,
\end{equation}
where $\nu_{\sb T}$ is an effective ''turbulent`` viscosity that needs to be estimated as a product of a typical length scale and a
typical velocity scale:
\begin{equation}
 \nu_{\sb T} \simeq l \sb{ch} v\sb {ch} \ .
 \label{scaling}
 \end{equation}
The typical velocity scale is estimated as a square root of the kinetic energy density per unit mass, $K(y)=\< |\~{\B v}(\B r)|^2\>/2$. The typical length scale in wall bounded flows is the distance to the wall, denoted here as $y$.
We thus end up with the estimate
\begin{equation}
\nu_{\sb T} \approx y \sqrt {K(y)}\ , \quad W(y)\approx y \sqrt {K(y)} S(y) \ ,
\end{equation}
up to constants of the order of unity.
%%%%%%%%%%%%%%%%%%%%%%%%%%%%%%%%%%%%%%%%%%%%%%%%%%%
\subsection{The case of quantum turbulence}

Consider next a superfluid flowing in the same channel configuration and explore the necessary modifications
of the classical arguments. For the inviscid component the viscosity is zero, and Eq.~(\ref{flux}) needs to be
modified, there is no explicit shear term in the momentum flux.
The Reynolds stress is however still there, but now correlating the fluctuations of the velocity of the superfluid.
We propose that Eqs.~(\ref{est}) and (\ref{scaling}) are still appropriate, except
that in the case of superfluids with random vortex tangles there is another characteristic length besides the distance to the wall. This length is the mean inter-vortex distance $\ell(y)$, which we propose to take as the estimate of $l\sb{ch}$ in \Eq{est}:
\begin{equation}\label{est1b}
l\sb {ch}\approx \ell(y)\,, \quad \ell(y) = 1/ \sqrt {\C L}(y)\ ,
\end{equation}
again up to constants of the order of unity.
Here ${\C L} (y)$ is the vortex line density  i.e. the vortex-line length per unit volume.

The characteristic velocity scale should be again provided by the square-root of a kinetic energy
density, but the appropriate one in this case is the kinetic energy density $K$ of the random vortex tangle.
To connect this energy density to the vortex line density we follow
Feinman\,\cite{Feynman} who related the kinetic energy density to the vortex line density as  $K \simeq \kappa^2\, \C L$.  For  quantum turbulence which is weakly varying  in space  at distances $\sim \ell$
it is tempting to assume that this estimate is satisfied locally:
  \begin{equation}\label{est1c}
K(y)\approx \kappa^2 \C L(y)\,, \ \Rightarrow \sqrt {K(y)}\approx   \kappa / \ell(y)\,,
\end{equation}
up to constants of the order of unity.
The upshot of this discussion is that for wall-bounded superfluid turbulence we can estimate the momentum flux as
  \begin{equation}\label{num}
 \Pi_{xy}(y)= \nu\sb m\, S(y)\,, \quad \nu\sb m\approx\, \kappa \ .
\end{equation}
 Here $\nu\sb m$  is an effective ''momentum`` kinematic viscosity and $S(y)$ is given by \Eq{flux}, in which now $V_x$ is the mean superfluid velocity in the stream-wise direction $x$,  ~$V_{x,\rm s }$. Henceforth we omit subscript $\rm s$ from the quantities related to the superfluid.

The conclusion is therefore as follows: the momentum flux towards the wall, which is carried by the superfluid turbulent velocity fluctuations, is determined by an effective momentum  viscosity of the order of the quantum circulation $\kappa$, and is independent of the vortex line density profile.

 Obviously, the derivation presented above is not rigorous, and it requires an experimental verification. Unfortunately, there are no experimental studies allowing us at present to clarify to which degree this simple prediction corresponds to physical reality.  We are therefore forced at the present time to test the prediction with numerical simulations.
%%%%%%%%%%%%%%%%%%%%%%%%%%%%%%%%%%%%%%%%%%%%%%%%%%%%

\section{Numerical simulations}
\label{numsim}
\subsection{Preparation of the system}

\begin{figure}[h]
   \includegraphics[scale=0.4]{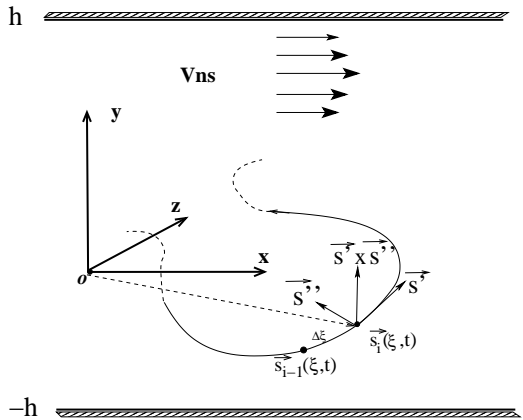}
   \caption{  \label{f:1} The Cartesian coordinates and the local vortex line  parametrization in a plane channel flow geometry. The counterflow velocity $ \B {V}\sb{ns}=\B {V}\sb n -\B {V} \sb s$ is oriented along the positive streamwise direction. }
   \end{figure}
%%%%%%%%%%%%%%%%%%%%%%%%%%%%%%%%%%%%%%%%%%%%%%%%%%%%%%%%%%%%%%

To test our ideas we performed numerical simulations of quantized vortex-line dynamics in a 3-dimensional planar channel geometry  $L_x\times L_y\times L_z$  of half-width $h$, $L_x=4h,L_y=2h,L_z=2h, h=0.05$~cm,  \Fig{f:1}.  To force turbulence we prescribed a time-independent profile of the stream-wise projection of the normal velocity $V_{x,{\rm n}}(y)\=V\sb n (y)$.  To find the
resulting vortex tangle configurations we used the vortex filament method~\cite{b:1}, taking into account the potential flow  to maintain the counterflow condition.

More details of the simulation method can be found in Refs.~\cite{b:Kond-1,Last}. Here we used the reconnection method \cite{Samuels92} and the line resolution $\Delta \xi=1.6\times 10^{-3}$ cm.  Periodic conditions were used in the streamwise and spanwise  directions.  Taking into account the fact that the boundary conditions for the superfluid component are still under discussion we adopted their simplest version: in the wall-normal $y$ direction    $V_{y}(\pm h)=0$ and ${\bm  s'(\pm h)}=(0,\pm 1,0)$ at the solid walls.

 Having selected a stationary profile of $ V\sb n(y)$ we initiated the simulations with a set of arbitrary oriented circular vortex rings and solved the equations for the vortex line evolution. Given the resulting dense vortex tangle  we found the profiles of all quantities of interest by averaging over periodic directions and over time for 20 seconds of steady-state evolution. To find   $W(y)$  we have calculated the superfluid velocity field $\B v(\B r,t)$  on a $128\times64\times128$ grid and then determined the profile $W(y)=\< v_x v_y \>-V_x V_y $.  Similarly the turbulent  kinetic energy profile was computed from  $2\,K(y)= \<  {\B v}^2 \> - V_x^2$.
%=======================FIG 2 ======================
\begin{figure}
  \includegraphics[scale=0.4 ]{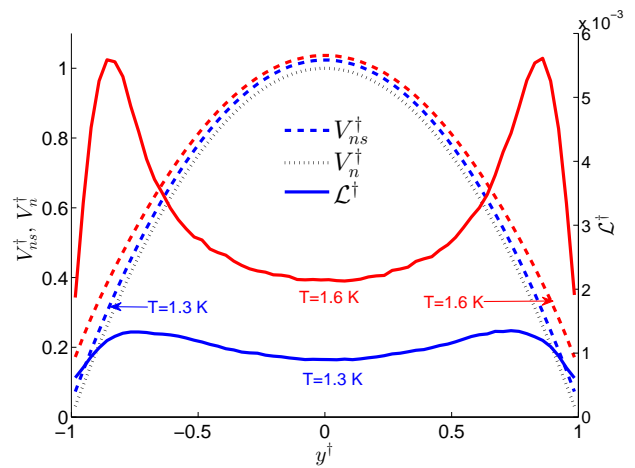}
 \caption{  \label{f:2}  Prescribed parabolic normal velocity profile $V \sb{n}^\dag(y^\dag)$ ( black dotted line), the resulting  profiles of the  counterflow velocity $V \sb{ns}^\dag(y^\dag)$ (dashed lines), and  the vortex line density $\C L^\dag(y^\dag)$ (solid lines). Blue lines show data for $T=1.3\,$K, red lines -- for $T=1.6\,$K. For both temperatures the centerline normal fluid velocity $V\sb n(0)=1.2$cm/s.}
\end{figure}

%===================== FIG 3 ========================
\begin{figure*}

\begin{tabular}{cc}
(a)& (b)\\
   \includegraphics[scale=0.4 ]{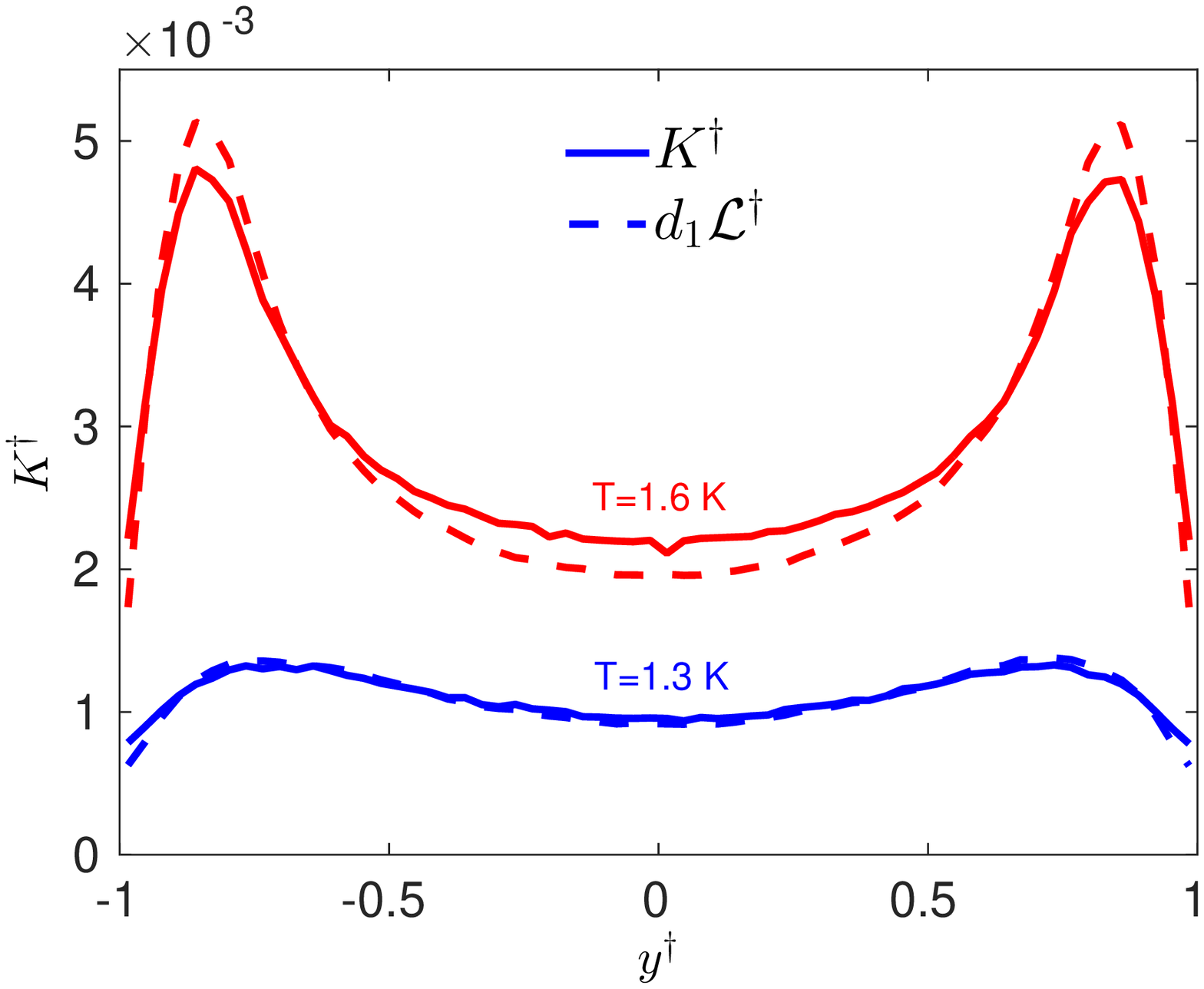}&
 \includegraphics[scale=0.4 ]{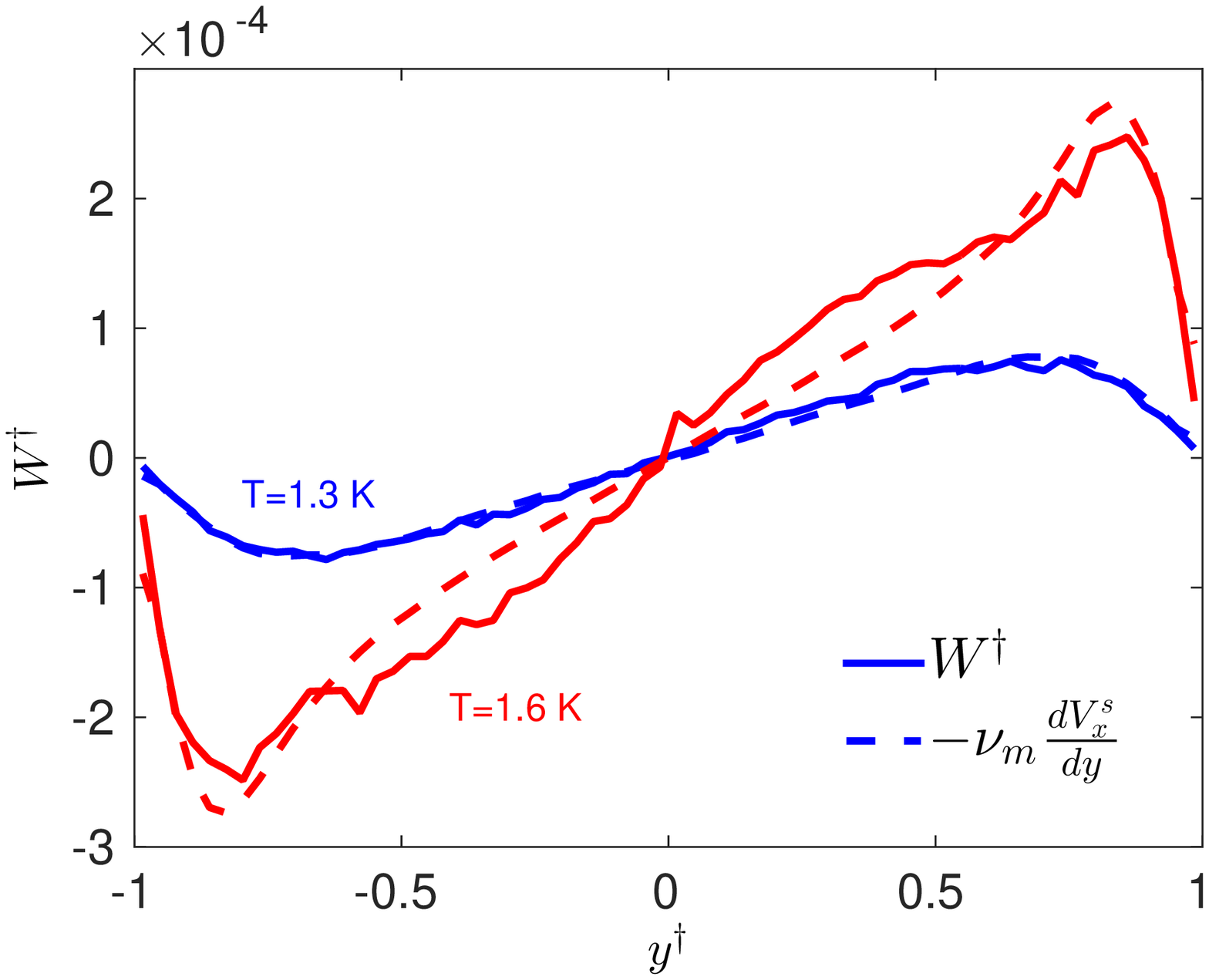}\\
\end{tabular}
\caption{  \label{f:3}  Color Online.  Verification of the closure for the  effective momentum viscosity: Panel (a): Comparison of the turbulent kinetic energy profiles $K^\dag(y)$ (solid  lines) with the vortex line density $d_1\C L^\dag(y)$( dashed line).  Here $d_1$ is a constant of the order of unity.  Panel~(b): Comparison of the Reynolds-stress profiles $W^\dag$ (solid lines), and the normalized shear profiles $- \nu\sb m S^\dag$ (dashed lines).   $T=1.3\,$K (blue lines) and $T=1.6\,$K (red lines). For both temperatures $V\sb{n}(0)=1.2$cm/s.}
\end{figure*}

\begin{figure}
 \includegraphics[scale=0.4]{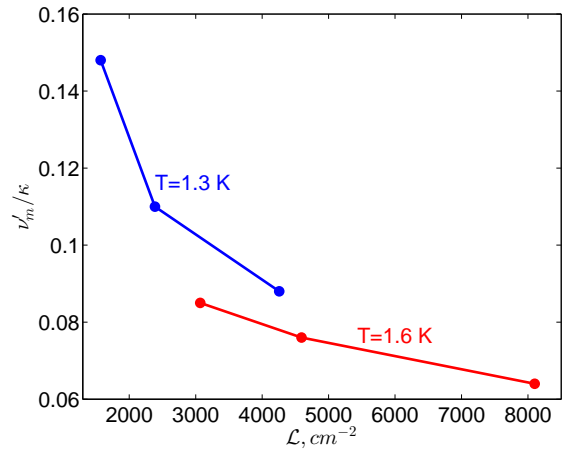}\\
  \caption{\label{f:4} Color online.  The temperature and the vortex line density dependence of the effective momentum viscosity $\nu\sb m/\kappa$.  }
\end{figure}

%=============================================
Before presenting the results of the simulations we stress that
the suggested closure for the Reynolds stress in the quantum turbulence case was based on an analogy with classical turbulence. It does not account for the two-fluid character of  superfluid $^4$He at intermediate temperatures. Therefore we cannot expect this closure to be optimal at temperatures for which the normal fluid and superfluid densities are comparable. Accordingly we focus here on the low temperature regime. For  $T=1.3\,$K and 1.6\,K the normal fluid densities are $\rho\sb n/\rho\approx 0.045$ and $0.162$, respectively~\cite{DB98}. A control test with  $T=1.9\,$K, when $\rho\sb n/\rho\approx 0.42$, will be shortly discussed later. At each temperature we tested three different values of the centerline counterflow velocity corresponding to the mean vortex line densities in the interval $(2-8)\cdot 10^3\,$cm$^{-2}$ (see \Fig{f:4}).
The results are presented in dimensionless units with the following normalizations:
 \begin{eqnarray} \nonumber
 &&y^\dag=  y/L\, ,V^\dag= V /V\sb n (0)\, , \C L^\dag(y) = \, \kappa^2 \C L/ [V\sb{n}(0)]^2\\\nonumber
 &&K^\dag(y) =2\, K(y)/ [V\sb{n}(0)]^2\, ,W^\dag (y)=W(y)/[V\sb{n}(0)]^2
\end{eqnarray}

 In Fig.\,\ref{f:2} we show by a black dotted line the prescribed parabolic profile of the normal velocity (identical for the two featured temperatures) and by blue and red dashed lines -- the resulting mean counterflow velocity profiles (for $T=1.3\,$K and 1.6\,K, respectively). At these  temperatures the normal-fluid density $\rho\sb n$ is much smaller than $\rho\sb s$. The condition of net mass-flow
  $\rho\sb n \< V\sb n(y)\>_y + \rho\sb s\< V (y)\>_y=0$  implies that  $ |\< V\sb n(y)\>_y| \gg  |\< V(y)\>_y|$.  Here $\< ...\>_y$ denotes averaging over the wall normal direction $y$.  Accordingly,  the  counterflow velocity profiles $V\sb{ns}(y)= V\sb n(y) - V(y)$ in \Fig{f:2} are not  very different from  the prescribed profile $V\sb n(y)$.

 Notice that the profiles of the vortex-line-density $\C L(y)$ and the counterflow velocity  $V\sb{ns}(y)$ are very different.  In fact, the vortex lines tend to concentrate  in regions with weaker counterflow velocity. Therefore the famous relation \cite{Vinen58} between the mean vortex line density and the mean counterflow velocity $\<\C L(y)\>^{1/2}_y = \gamma (\<V\sb{ns}(y)\>_y-v_0$) with constant coefficients $\gamma$ and $v_0$ holds only globally and is not fulfilled locally across the channel.

\subsection{Confirmation of the concept of $\bm{\nu\sb m}$}

To test (and confirm) the suggested closure\,\eqref{num}  we demonstrate in \Fig{f:3}(a) that the profile of the turbulent kinetic energy $K(y)$ is indeed proportional to the vortex line density profile, as required by  \Eq{est1c} that leads to the closure\,\eqref{num}.  Indeed, there is an excellent  correspondence between $K(y)$ and $\C L(y)$ profiles for $T=1.3\,$K (blue lines) with only a slight discrepancy between them for $T=1.6\,$K (red lines). Even for $T=1.9\,$K (not shown) the agreement between $K(y)$ and $\C L(y)$ is fairly good.

 Next, we compared in \Fig{f:3}(b) the profiles of the Reynolds stress $W^\dag(y)$ with the normalized shear profile $[\nu\sb m S]^\dag (y) = - \nu\sb m [d V(y)/d y] /[V\sb{n}(0)]^2$.
The closure approximation $[\nu\sb m S]^\dag(y^\dag)$ follows closely the behavior of  $W^\dag(y^\dag)$: notice the (almost) linear profile in the center of channel and the drop near the wall.
As for the kinetic energy, there is an excellent  correspondence between $W^\dag(y^\dag)$ and $[\nu\sb m S]^\dag(y\dag)$ profiles for $T=1.3\,$K (blue lines) and some slight deviations between them for $T=1.6\,$K (red lines).
For a higher temperature $T=1.9\,$K (not shown) the closure $[\nu\sb m S]^\dag(y^\dag)$ reproduces $W^\dag(y^\dag)$ only qualitatively. We conjecture that this is due to the influence of the normal fluid component which is not accounted for in the closure.

The comparison of the numerically found profiles $W(y)$ and $\nu\sb m S(y)$ allows us to measure the effective momentum viscosity   $\nu_m$.   Its    temperature and mean-vortex line density $\< \C L \>$ dependence  is shown in \Fig{f:5}. The viscosity $\nu\sb m(T,\< \C L\>)$ decreases upon increasing either the temperature or the vortex line density. Notice that the temperature dependence of $\nu\sb m(T)$ is opposite to that of the Vinen effective energy viscosity $\nu\sb e(T)$, which increases with temperature.

\section{The nature of the vortex  tension force}
\label{tense}
\subsection{Analysis}

The idea of taking into account the effect of the small scale structures of the quantum vortex tangle on the macroscopic  equations of motion is not new. In the so-called Hall-Vinen-Bekarevich-Khalatnikov (HVBK) "coarse-grained"
equations \cite{HallVinen56,BK61} for the large scale normal and superfluid velocities $\B v\sb n(\B r,t)$ and $\B v(\B r,t)$ it leads to the notion of ``vortex tension force"
\begin{equation}\label{tension}
\BC T= - \kappa \frac{\Lambda}{4\pi}\,  \B \omega \times(\B \nabla\times \B {\hat \omega})\,,   \quad  \Lambda =\ln \frac{\ell}{a_0}\ .
\end{equation}
Here the superfluid vorticity  $\B \omega=(\B \nabla \times \B v)$ and $\B {\hat \omega}= \B \omega /|\B \omega|$. This
force appears in the equation of motion of $\B v(\B r,t)$.

In this section we present an analytic calculation of the gradient of the Reynolds stress tensor, $(\B {\~v} \cdot \B \nabla) \B {\~v}$ under the simplifying assumption that the fluctuating part of the velocity is produced by a local velocity close to the  vortex lines. We will show that the result coincides with the tension force $\BC T$.

To this end consider first the velocity close to the vortex line \cite{Saffman}:
\begin{eqnarray}\nn
 \bm { \~{v}(\bm R)}&=&\frac{\kappa}{2\pi { R}^2}\big [  X (\B s' \times \B s'')  -Y \B s''\big]\Big (\frac 1 {|\B s''|} + \frac {X }{2}\Big) \\&&
+\frac{\kappa (\Lambda-1)}{4\pi} \B s'\times \B s'' \ .\label{vel}
\end{eqnarray}
Here $\B s(\xi)$ is the radius vector to the vortex line parameterized by the arc length $\xi$, see \Fig{f:1}.  The derivative $\B s'=d \B s/d \xi$ is a unit vector in the direction of the vortex line,  $\B s''= d^2 \B s/ d \xi^2$ is a vector normal to the vortex line with absolute value equal to its local curvature and $\bm s'\times \bm s''$ is the bi-normal vector whose absolute value is also equal to the curvature. The coordinate   ${\bm R}$ is the  radius vector from the point $\B s(\xi)$ on the vortex line to a point close to the vortex line. We denote by $X $ the projection  of $\bm{{R}}$ on the  direction of $\bm s''$ and by $Y$  the projection on the  direction of $\bm s' \times \bm s''$. The first term on the right-hand-side of \Eq{vel} is the velocity of a rectilinear vortex line, the second one takes into account distortions in the axial symmetry due to curvature; the  last term is the local self-induced velocity.

In the Appendix we show that
substituting \Eq{vel} into $(\B {\~{v}}\cdot\B \nabla)\B {\~v}$ and integrating over a cylindrical volume around the vortex line we get:
\begin{equation}\label{res}
\<(\B {\~v}\cdot\B \nabla)\B {\~v}\>=-\frac{\kappa^2\Lambda }{8\pi } \  \< \B s'' \> \C L \ .
\end{equation}

 Next we will show that  \Eq{res} is equivalent to \Eq{tension} for $\BC T$. Note first that
 $  \< \bm s''\>=  \< (\bm s' \cdot \nabla)\bm s'\>=  \<\nabla ( |\bm s'|^2/2)-\bm s' \times (\nabla \times \bm s')\>$.
 Now we  can rewrite \Eq{res} as follows:
 \begin{equation}\label{Eq12}
\<(\B {\~v}\cdot\B \nabla)\B {\~v}\>= \frac{\kappa^2\Lambda }{8\pi } \C L \<\bm s'\times(\nabla \times \bm s') \> \ .
\end{equation}

At this point one can already see that equation \Eq{Eq12} and \Eq{tension} have the same structure up to a change of $\bm s'$ by $\bm{\hat\omega}$ and $\kappa \C L \bm s' $ by $\bm{\omega}$, required in the derivation of the HVBK equations \cite{HallVinen56,BK61}.

The upshot of this calculation is that the gradient of the Reynolds stress tensor of velocity fluctuations $\<(\~{\B v}\cdot\B \nabla)\~{\B v}\>$ on the scale of the inter vortex distance and  the
tension force $\BC T$, given by \Eq{tension} are actually the same.
The tension force was derived and used in the context of quantum turbulence; the Reynolds stress tensor is more familiar in the context of classical turbulence. We can summarize the consequences of this identity, using \Eqs{num} and \eqref{res} to provide   a remarkably simple  representation of the complicated vector structure\,\eq{tension} of the tension force via the mean superfluid velocity $\B V(\B r)$:
 \begin{equation}\label{force}
 \BC T(\B r)=  \nu\sb m \Delta \B V(\B r)\ .
 \end{equation}

We believe that this new interpretation carries an important advantage.
In the original \Eq{tension} for $\BC T$ the vorticity $\B \omega$ is dominated by small scale ($\sim \ell$) fluctuations.  On these small scales the validity of the HVBK equations is questionable.  On the other hand, in the closure approximation\,\eqref{force} the main contribution to the mean velocity $\B V(\B r)$ comes from the largest scales in the flow, for with  course-grained HVBK equations are quite acceptable.
%%%%%%%%%%%%%%%%%%%
\subsection{Numerical confirmation}

\begin{figure}
 \includegraphics[scale=0.4]{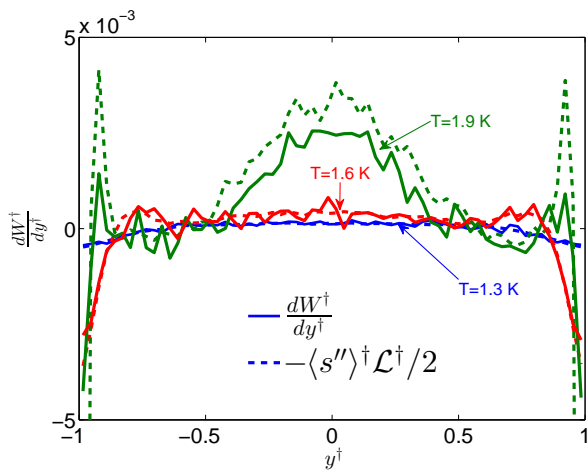}
  \caption{\label{f:5} Color online. Comparison of the numerical results for  the gradient of the Reynolds stress $ d W^\dag(y^\dag)/ dy^\dag $ (solid   lines) and the tension force \eqref{res} (dashed   lines) for   at $T=1.3\,$K,  $T=1.6\,$K and $T=1.9\,$K. $V\sb{n}(0)=1.2$cm/s.}
\end{figure}

Eq.~(\ref{res}) was derived on the basis of the approximation that the fluctuating part of velocity is given by the local velocity in the proximity of the vortex line. Larger scale fluctuations were neglected. We therefore need to
test the approximation using our numerics.
The justification of the approximation is shown in \Fig{f:5}, where numerical results  for $ d W^\dag / dy^\dag $ and  $\< |s''|\> \C L^\dag/2 $  (see Eq.~\eqref{res})  are  plotted as solid and dashed  lines respectively.  Clearly, these lines practically coincide (up to numerical noise), without any fitting parameters, for all the studied temperatures, including $T=1.9\,$K.

We conclude that our analytical and numerical results demonstrate that  \emph{the gradient of the Reynolds stress and the tension force, both  originated from the motions of the random vortex tangle, are identical objects.} \vskip 0.2cm

\section{Summary and the road ahead}
\label{summary}

In superfluid turbulence there are two related phenomena that stem from the existence of a tangle of
quantized vortexes. Although the intervortex distance $\ell$ is small, the effect of the vortex tangle appears
also in the coarse-grained dynamics.  One effect is the Vinen effective viscosity $\nu'$, that describes the rate of energy dissipation. In this paper
we introduced a second effective viscosity $\nu\sb m(T)$ that describes the transfer of linear mechanical momentum toward regions with lower mean velocity. It appears in the friction force $\nu\sb m(T)\Delta \B V(\B r,t)$. This friction force operates between layers with different superfluid velocities $\B V(\B r,t)$, mediated by the random vortex tangle.

The second result of this paper follows from an analysis of the gradient of the vortex tangle-induced Reynolds stress tensor and of the vortex-tension force $\BC T$. We show that these two objects are just different names for the same mechanism of transfer of linear mechanical momentum.

 Additional analysis is required to fully understand the properties of the effective momentum viscosity $\nu\sb m(T)$ and to clarify its range of applicability. These subjects go beyond the scope of this paper and will be dealt with in future publications.

%========================== FIG 4 ==============

%========================== FIG 5 ==============

\vskip 0.2cm
%========================== ==============

\section{Appendix: Calculation of the tension force}
In order to calculate  $(\bm{\~v}\cdot\bm{\nabla})\bm{\~v}$
we  start by writing explicitly the contributions to the velocity close to the vortex line given by \Eq{vel}:

\begin{eqnarray}\label{vGradv}
v_{\sb X}&=&\frac{\kappa}{4\pi}\left[-\frac{2 Y}{{R}^2}-\frac{X Y}{ {R}^2}|s''|\right]\, , \\\nonumber
v_{\sb Y}&=&\frac{\kappa}{4\pi}\left[\frac{2 X}{{R}^2} -\frac{Y^2}{ {R}^2}|s''| +|s''|\Lambda\right]\, .
\end{eqnarray}
These expressions were obtained as a consistent first order expansion in curvature $|s''| $.

Next, we calculate the derivatives of the velocities with respect to local coordinates $X$ and $Y$:
\begin{eqnarray}\label{vGradv2}
\frac{\partial v_{\sb X}}{\partial X}& =& \frac{\kappa}{2\pi}\left[\frac{2 X Y}{ {R}^4} -\frac{Y}{2 {R}^2}|s''|+\frac{X^2 Y}{ {R}^4}|s''|\right]\, , \\\nonumber
\frac{\partial v_{\sb X}}{\partial Y} &=&\frac{\kappa}{2\pi}\left[ -\frac{1}{ {R}^2}+\frac{2{Y}^2}{ {R}^4} -\frac{X}{2  {R}^2}|s''|+\frac{X {Y}^2}{  {R}^4}|s''|\right]\, ,
\end{eqnarray}
\begin{eqnarray*}
\frac{\partial v_{\sb Y}}{\partial X} &=& \frac{\kappa}{2\pi}\left[\frac{1}{ {R}^2}-\frac{2{X}^2}{ {R}^4}+\frac{{Y}^2 X}{  {R}^4}|s''| -\frac{X}{2  {R}^2}|s''|\right]\, ,\\\nonumber
\frac{\partial v_{\sb Y}}{\partial Y}&=&\frac{\kappa}{2\pi}\left[ -\frac{2 Y X}{ {R}^4} -\frac{Y}{ {R}^2}|s''|+\frac{{Y}^3}{ {R}^4}|s''| -\frac{Y}{2 {R}^2}|s''|\right]\, .
\end{eqnarray*}

To average $(\bm{\~v}\cdot\bm{\nabla})\bm{\~v}$ in a cylindrical volume around the vortex line we use curvy-linear coordinate system where the first coordinate is the arclength $\xi$ and the other two are the polar coordinates  in the plane perpendicular to $\bm s'$ with a polar axis that form an angle $\psi(\xi)$ with $\bm s''$. Here $\psi(\xi)$ defined via torsion of the vortex line $\tau$ as $\tau=d\psi(\xi)/d\xi$.

The coordinate system, defined in this way, is orthogonal with metric coefficients\cite{Saffman}:
\begin{equation}
h_{\theta}=R; \, h_{R}=1; \, h_{\xi}=1-R|s''|\cos(\theta-\psi)\, ,
\end{equation}
where $\theta$ is an azimuthal angle in the polar plane.

Then the relation between Cartesian and polar coordinates is given by
$X=R\cos(\theta-\psi), \, Y=R\sin(\theta-\psi)$,
such that the element of volume is:
\begin{equation}
 d\Omega=R(1-R|s''|\cos(\theta-\psi))dR d\theta d\xi \, .
\end{equation}

Now, substituting Eqs.\eqref{vGradv} and \eqref{vGradv2} into $(\bm{\~v}\cdot\bm{\nabla})\bm{\~v}$ and integrating in the cylindrical volume around vortex line, after some algebra we get:
\begin{equation}
\frac{1}{\Omega}\int(\bm{\~v}\cdot\bm{\nabla})\bm{\~v}d\Omega=-\frac{\kappa|s''|}{8\pi \Omega}\int d\xi \int\limits_{a_0}^{\ell}\frac{dR}{R}=-\frac{\kappa \<\bm{s''}\>}{8\pi}\Lambda \C L
\end{equation}

% \noindent{\bf  Acknowledgements} --  bla


\begin{thebibliography}{99}
\bibitem{DB98}  R. J. Donnelly, C. F. Barenghi , The Observed Properties of Liquid Helium at the Saturated Vapor Pressure,
 J. Phys. Chem. Ref. Data \textbf{27,} 1217(1998).

\bibitem{Donnely}
R. J. Donnelly, Quantized Vortices in Hellium II (Cambridge
3 University Press, Cambridge, 1991).

\bibitem{Feynman} R. P.Feynman,
 %"Application of quantum mechanics to liquid helium".
  Progress in Low Temperature Physics \textbf{1}: 17–53, (1955).

\bibitem{Vinen} W. F. Vinen and J. J. Niemela, J. Low Temp. Phys. {\bf 128}, 167 (2002).


\bibitem{HallVinen56} Hall HE, Vinen WF (1956)
%The rotation of liquid helium II. II. The theory of mutual friction in uniformly rotating helium II. Proc R Soc Lond A Math
Phys Sci {\bf 238} 215–234.
\bibitem{Vinen58}W. F. Vinen Proc. R. Soc. Lond. A 1958 {\bf 243}, 400-413

\bibitem{BK61}I.L. Bekarevich, I.M. Khalatnikov,
%Phenomenological derivation of the equations of vortex motion in He II,
 Sov. Phys. JETP \textbf{13} (3), 643-646 (1961).

\bibitem{Frisch} U. Frisch. Turbulence: the legacy of A.N. Kolmogorov
Cambridge university press.

\bibitem{Pope} S.B. Pope, Turbulent flows, Cambridge university press.


\bibitem{Vinen2} W. F. Vinen, J. Low Temp. Phys. 161, 419 (2010).

\bibitem{SS} L. Skrbek and K. R. Sreenivasan, Phys. of Fluids, \textbf{24}, 011301 (2012)

\bibitem{Nemir} S. Nemirovskii, Physics Report \textbf{524} 85 (2013).

\bibitem{WIS-2015} L. Boue, V S. L'vov., Y. Nagar., S. V. Nazarenko, A. Pomyalov., and I. Procaccia. Energy and Vorticity Spectra in Turbulent Superfluid 4He from T = 0 to T-lambda. Phys. Rev. B 91, 144501 (2015).

\bibitem{2} J. Boussinesq, Theorie de l\`{a}coulement tourbillant. (Mem.
Pres. Acad. Sci., Paris, 1877), Vol. XXIII, p. 46.

\bibitem{1} L. Prandtl, Z. Angew. Math.Mech., \textbf{5} 136 (1925).

\bibitem{b:1}  K.W. Schwarz, Three-dimensional vortex dynamics in superfluid $^4$He: Homogeneous superfluid turbulence,
 Phys. Rev. B  \textbf{38,}  2398 (1988).

\bibitem{b:Kond-1} L. Kondaurova,   V. L'vov,  A. Pomyalov  and I. Procaccia, Structure of a quantum vortex tangle in $^4$He counterflow turbulence,
Phys. Rev. B   \textbf{89,} 014502 (2014).
\bibitem{Last} D. Khomenko, L. Kondaurova, V.S. L'vov, P. Mishra, A. Pomyalov and I.Procaccia, PRB 91, 180504(R), 2015

\bibitem{Samuels92}  D. C. Samuels,  Velocity matching and Poiseuille pipe flow of superfluid helium,
Phys. Rev. B {\bf 46,} 11714 (1992).

\bibitem{Jou} D. Jou and M.S. Mongiovi PRB  74  054509,(2006)

\bibitem{14} R. M. Ostermeyer and W. I. Glaberson, J. Low Temp.
Phys. \textbf{21}, 191 (1975).

 \bibitem{Saffman} P.G. Saffman, Vortex dynamics (Cambridge University Press, 1992)


\end{thebibliography}
\end{document}